\documentclass[prb,twocolumn,showpacs,amsmath,amssymb,citeautoscript]{revtex4}
\usepackage{epsfig}
\usepackage{amsmath}
\usepackage{amssymb}
\usepackage{graphicx}
\usepackage{multirow}
\usepackage[usenames]{color}
\usepackage[english]{babel}

\begin{document}

\title{Magnetostructural Effect in the Multiferroic BiFeO$_3$-BiMnO$_3$ Checkerboard from First Principles} 
\author{L. P\'alov\'a, P. Chandra and K. M. Rabe}
\affiliation{Department of Physics and Astronomy, Rutgers University, 
Piscataway, NJ 08854}
\date{\today}

\begin{abstract}
Using first principles calculations, we present a magnetostructural effect in
the BiFeO$_3$-BiMnO$_3$ nanocheckerboard that is not found in either bulk
parent compound or in
BiFeO$_3$-BiMnO$_3$ superlattices.
We also demonstrate that the atomic-scale checkerboard has a multiferroic
ground state with the desired properties of each constituent material:
polar and ferrimagnetic due to
BiFeO$_3$ and BiMnO$_3$ 
respectively.
\end{abstract}


\maketitle

There is currently tremendous interest in finding new multiferroic (ferroelectric and ferromagnetic) materials 
with large magnetoelectric coupling. Advances in the synthesis of artificially structured materials have stimulated 
efforts to design new multiferroic heterostructures, with first principles methods being an essential tool for the 
identification and investigation of promising systems. 
In this Letter, we report the first-principles identification and characterization of an unusual heterostructure, a multiferroic atomic-scale 2D nanocheckerboard \cite{Zheng04,Yeo06,Guiton07,Driscoll08} of BiFeO$_3$-BiMnO$_3$, with properties that critically depend on the geometry and are not present 
in either bulk or layered structures of the constituent materials. In particular, the 2D checkerboard geometry 
leads to magnetic frustration and to quasi-degenerate magnetic states that can be tuned by an external perturbation 
that changes the crystal structure, such as an electric field. 
This results in a novel magnetostructural effect, adding to previous examples of magnetostructural coupling such as bulk\cite{Kozlenko03} and layered\cite{Murata07} manganites,
epitaxial $EuTiO_3$\cite{Fennie06} and $EuSe/PbSe_{1-x}Te_x$ 
multilayers.\cite{Lechner05}

Our first principles calculations are performed using density functional theory within the 
local spin-density approximation (LSDA)+U method as implemented in 
the Vienna Ab-initio Simulation Package VASP-4.6.34~\cite{Kresse93}.
We test the robustness of our results with two different implementations of the rotationally invariant LSDA+U version, the first as introduced by Liechtenstein~\cite{Liechtenstein95} with $U_{Fe}=U_{Mn}=5 eV$, $J_{Fe}=J_{Mn}=1 eV$, and the second due to
Dudarev~\cite{Dudarev98}, with $U_{Mn}^{eff}=5.2 eV$, $U_{Fe}^{eff}=4 eV$, where $U^{eff}=U-J$.
It has been shown that these $U$ and $J$ values 
match experimental data in bulk BiFeO$_3$~\cite{Neaton05}; the
value $U^{eff} =5.2 eV$ has been used for previous bulk BiMnO$_3$
ground state calculations~\cite{Baettig07}.
We use projector-augmented wave potentials (PAW)~\cite{Blochl94,Kresse99}
and treat explicitly 15 valence electrons for Bi ($5d^{10}6s^2 6p^3$),
14 for Fe ($3p^6 3d^6 4s^2$), 13 for Mn ($3p^6 3d^5 4s^2$),
and 6 for O ($2s^2 2p^4$).
The cutoff energies for the plane wave basis set are $550 eV$and $800 eV$ 
in the ionic relaxations and for subsequent self-consistent energy 
calculations respectively.
Gaussian 
broadening of the partial occupancies for each wavefunction is $0.05 eV$.
A Monkhorst-Pack k-point grid~\cite{Monkhorst76} 
is generated with density $4\times4\times4$
for $(\sqrt{2}\times\sqrt{2}\times 1) a_0$ double perovskite 
and $4\times4\times2$ for $(\sqrt{2}\times\sqrt{2}\times 2) a_0$ four perovskite cells.
Ions are relaxed towards
equilibrium positions until the Hellmann-Feynman forces are less than $10^{-3} eV/\AA$.
The spontaneous polarization is calculated by the Berry phase method~\cite{King-Smith93}
with k-point mesh twice as dense as in the energy calculations.

\begin{figure} 
\begin{center}
\includegraphics[scale=0.4]{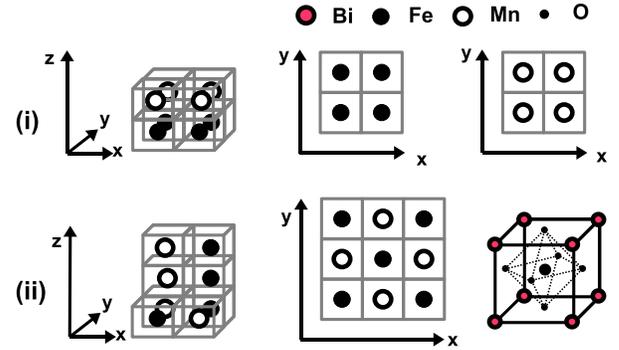}
\caption[checkerboard]
        {
        (i) BiFeO$_3$-BiMnO$_3$ superlattice with alternation of Fe/Mn planes.
        (ii) (left) BiFeO$_3$-BiMnO$_3$ checkerboard.
        Checkerboard ordering of Fe/Mn atoms in the ($xy$) plane,
        pillars of the same composition form along the $z$-direction.
        (right) Ideal perovskite unit cell. Perovskite cells with Fe/Mn atoms on the B-site
        repeat according to the checkerboard pattern (ii), or layered geometry (i).}
\label{checkerboardfig}
\end{center}
\end{figure}

\begin{figure}
\begin{center}
\includegraphics[scale=0.33]{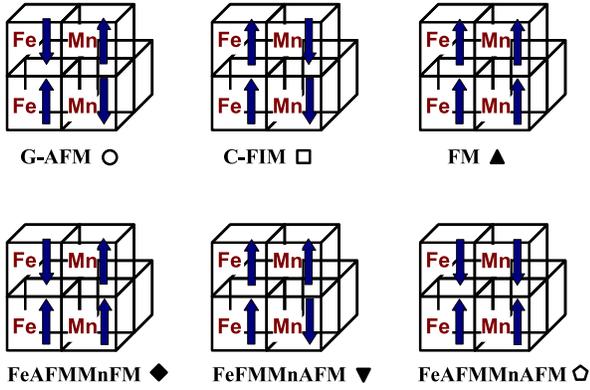}
\caption[magtype]
{
From top left to bottom right:
(i) G-AFM: rocksalt type antiferromagnetic (AFM) order,
(ii) C-FIM: AFM order in horizontal planes,
ferromagnetic (FM) order along Fe/Mn pillars,
(iii) FM order,
(iv) FeAFMMnFM: AFM order along Fe pillars, FM order along Mn pillars,
(v) FeFMMnAFM: FM order along Fe pillars, AFM order along Mn pillars,
(vi) FeAFMMnAFM: AFM order along Fe/Mn pillars, but FM order in horizontal planes.
}
\label{magtypefig}
\end{center}
\end{figure}

We consider four formula units (perovskite cells),
two each with Fe and Mn atoms on the $B$-site,
which we repeat periodically in space.
For the planar checkerboard, we alternate iron (Fe) and manganese (Mn) atoms at the atomic level 
to form pillars of the 
same composition as in Fig.~\ref{checkerboardfig}(ii). 
For the layered superlattice, we alternate single unit cell layers along z, as in Fig.~\ref{checkerboardfig}(i). In both cases, the supercell is $\sqrt{2}a_0\times\sqrt{2}a_0\times 2 a_0$.

We study various
collinear spin orderings of the magnetic
Fe and Mn atoms, shown for the checkerboard in Fig.~\ref{magtypefig}.
FeFM and FeAFM refer to ferromagnetic and antiferromagnetic ordering
respectively for the Fe moments in the relevant structural component 
(pillar for the checkerboard, layer for the superlattice); similarly 
MnFM and MnAFM describe the spin ordering of the Mn moments. 
In the checkerboard, this notation fully specifies the states considered. 
For the superlattice, the remaining ambiguity is resolved as follows:
FeAFMMnAFM magnetic order designates AFM ordered Fe and Mn planes
with FM order along the mixed Fe-Mn chains in the $z$ direction, while G-AFM designates the case with AFM order along the mixed chains; similarly, FeFMMnFM designates FM ordered Fe and Mn planes with AFM order along the mixed Fe-Mn chains in the $z$ direction, while FM designates the case with FM order along the mixed chains.

In searching for the ground state crystal structure for each spin ordering, we consider structures generated by three typically unstable modes 
of the cubic perovskite structure:~\cite{Stokes02} 
the zone center polar mode $\Gamma_4^-$, the M$_3^+$ oxygen octahedron rotations (all rotations about a given axis in phase) and R$_4^+$ rotations (sense of rotations alternates along the rotation axis). 
We freeze in selected modes and combinations of modes and optimize atomic displacements and lattice parameters 
in the resulting space groups.

\begin{table*} 
\begin{center}
\caption{Calculated magnetic energies in an ideal perovskite setting with lattice constant $a_0=3.893\AA$
for various magnetic states in the checkerboard,
layered superlattice, and bulk BiFeO$_3$ and BiMnO$_3$.
Value of $U=5eV$ and $J=1eV$ is used.
} 
\label{tablemagenergiesU4}
\begin{tabular}{|cc|cc|cc|cc|}
\hline
 checkerboard & & layered superlattice & & BiFeO$_3$ & & BiMnO$_3$ & \\
 magnetic state & $\Delta E[eV/f.u.]$ & magnetic state & $\Delta E[eV/f.u.]$ & mag. state &  $\Delta E[eV/f.u.]$ & mag. state & $\Delta E[eV/f.u.]$ \\
\hline
FeAFMMnFM & 0.000   & FeAFMMnFM  & 0.000   & - & & - & \\
\hline
FM        & 0.022 & FM         & 0.111 & FM    & 0.360 & FM    & 0.000   \\
\hline
C-FIM     & 0.076 & FeFMMnFM   & 0.136 & C-AFM & 0.115 & C-AFM & 0.293 \\
\hline
FeAFMMnAFM& 0.081 & FeAFMMnAFM & 0.135 & A-AFM & 0.223 & A-AFM & 0.116 \\
\hline
G-AFM     & 0.114 & G-AFM      & 0.181 & G-AFM & 0.000   & G-AFM & 0.494 \\
\hline
FeFMMnAFM & 0.119 & FeFMMnAFM  & 0.260 & - & & - & \\
\hline
\end{tabular}
\end{center}
\end{table*}

First, to investigate 
the effect of $B$-site cation geometry on the magnetic ordering,
we present results in Table~\ref{tablemagenergiesU4} for the magnetic ordering energies when the structures are held fixed to the ideal perovskite reference structure.
In the layered superlattice and both bulk parent systems, 
the difference in energy between magnetic ground state
(FeAFMMnFM in the superlattice, G-AFM in bulk BiFeO$_3$ and FM in bulk BiMnO$_3$)
and the first alternative state is 
$0.11-0.12 eV/f.u.$;
this difference corresponds to 
a relatively large energy and we do not expect a transition
to a different magnetic state.
The highest energy magnetic states are more than $0.26 eV/f.u.$ apart.
On the other hand in the checkerboard,
all magnetic states are found quasi-degenerate and 
are confined within the energetical window of $0.12 eV/f.u.$, 
that is, all are lower than the lowest states in the layered superlattice and bulk parent compounds.
Indeed, the closest magnetic state to the FeAFMMnFM ground state is now only $0.022 eV/f.u.$ 
higher, making it much more plausible for a magnetic transition to occur.

\begin{figure} [t!]
\begin{center}
\includegraphics[scale=0.35,angle=-90]{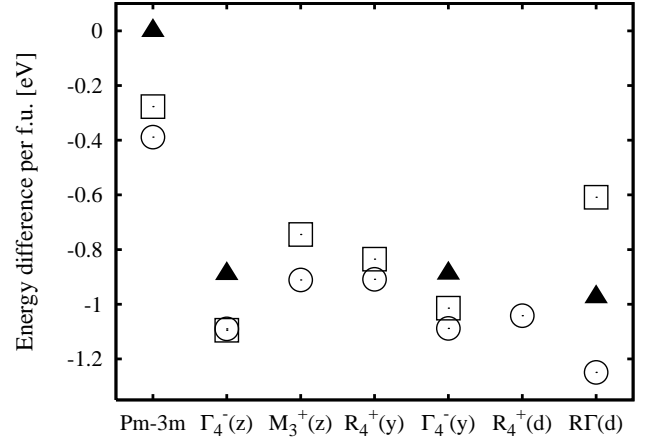}
\caption[ediffdistbfo]
{Structural energetics of bulk BiFeO$_3$. 
Energy difference per perovskite cell (f.u.) 
for different magnetic orderings (see Fig.~\ref{magtypefig}) and 
structural distortions:
(1) $Pm\bar{3}m$: no distortion - ideal perovskite,
(2) $\Gamma_4^- (z)$:  polar distortion along $z$ axis, 
(3) $M_3^+ (z)$: $+$ oxygen octahedra tilts about $z$ axis, 
(4) $R_4^+ (y)$: $-$ oxygen octahedra tilts about $y$ axis, 
(5) $R_4^+ (y)$ and $\Gamma_4^- (y)$ ($R\Gamma (y)$): linear combination of (4) and (2) along $y$ axis,
relaxes back to polar $\Gamma_4^- (y)$ with zero tilting angle,
(6) $R_4^+ ([111])$ ($R_4^+(d)$): $-$ oxygen octahedra tilts about $[111]$ axis,
(7) $R_4^+ ([111])$ and $\Gamma_4^- ([111])$ ($R\Gamma(d)$): linear combination of (6) and (2) 
along $[111]$ ($d$),
where $d$ refers to the cube diagonal direction.
}
\label{ediffdistbfofig}
\end{center}
\end{figure}

Next we study the energetics of the structural distortion and 
its effect on the spin order.
Before discussing results for the BiFeO$_3$-BiMnO$_3$ checkerboard,
we look at the structural energetics of the two bulk constituent 
materials, BiFeO$_3$ and BiMnO$_3$.
We plot energies for various magnetic orderings in seven types of structural distortions of bulk
BiFeO$_3$ in Fig.~\ref{ediffdistbfofig}.
Our calculation verifies the R3c ground state of BiFeO$_3$: counter-rotations of the oxygen octahedra ($R_4^+$) 
and polar ionic distortions ($\Gamma_4^-$) along the $[111]$ axis 
are most energetically favorable.~\cite{Michel69,Neaton05}
The ground state structure has G-type AFM order and
spontaneous polarization $90 \mu C/cm^2$ along [111] axis.
For all structural distortions considered, the lowest energy magnetic ordering is G-AFM.

\begin{figure} [t!]
\begin{center}
\includegraphics[scale=0.35,angle=-90]{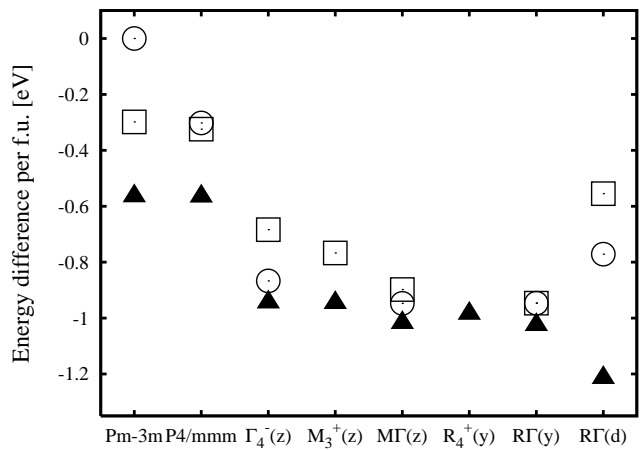}
\caption[ediffdistbmo]
{
Structural energetics of bulk BiMnO$_3$. 
Energy difference per perovskite cell (f.u.) 
for various structural distortions (see Fig.~\ref{ediffdistbfofig})
and magnetic orderings (see Fig.~\ref{magtypefig});
$P4/mmm$ corresponds to a tetragonally distorted perovskite cell with ideally positioned atoms
and $M\Gamma (z)$ is a linear combination of 
rotational $M_3^+ (z)$ and polar $\Gamma_4^- (z)$ modes.
}
\label{ediffdistbmofig}
\end{center}
\end{figure}

We study the structural energetics of bulk BiMnO$_3$ in a similar way;
the plotted energies for various structural distortions and magnetic orderings are
presented in Fig.~\ref{ediffdistbmofig}.
We find the lowest energy structure with R3c symmetry,
the same as the ground state of BiFeO$_3$.
The lowest energy structure 
has a half-metallic character
and is ferromagnetic.
This structure is not the  
monoclinic centrosymmetric ground state $C2/c$ of bulk BiMnO$_3$
which has a larger unit cell than that considered here.~\cite{Belik06}
However our calculation shows that it lies close to the ground state (only $43$ $meV/f.u.$ above the GS).
For all structural distortions considered, the lowest energy magnetic ordering is FM.

In the layered BiFeO$_3$-BiMnO$_3$ superlattice,
we calculate magnetic energies for the rocksalt type G-AFM and FeAFMMnFM layered magnetic states
in two structural distortions.
For $R_4^+ (y)\&\Gamma_4^-(y)$, we find 
$\Delta E$ $=$ $-0.504 eV/f.u.$ for G-AFM and 
$\Delta E$ $=$ $-0.553 eV/f.u.$ for FeAFMMnFM 
with respect to the FeAFMMnFM magnetic state in the ideal perovskite cell (see Table~\ref{tablemagenergiesU4}).
For $R_4^+ ([111])\&\Gamma_4^-([111])$, we find 
$\Delta E$ $=$ $-0.752 eV/f.u.$ for G-AFM and 
$\Delta E$ $=$ $-0.761 eV/f.u.$ for FeAFMMnFM.
For both structural distortions considered, the lowest energy magnetic ordering is FeAFMMnFM.

\begin{figure} [t!]
\begin{center}
\includegraphics[scale=0.35,angle=-90]{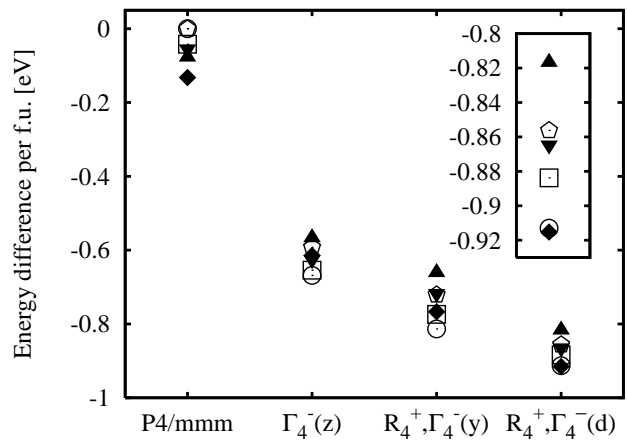}
\caption[ediffdistbfmo]
{
Structural energetics of BiFeO$_3$-BiMnO$_3$ checkerboard. 
Energy difference per perovskite cell (f.u.) 
for different magnetic orderings (see Fig.~\ref{magtypefig}) and 
structural distortions:
(1) $P4/mmm$,
(2) $\Gamma_4^- (z)$,
(3) $R_4^+ (y)$ and $\Gamma_4^- (y)$,
(4) $R_4^+ ([111])$ and $\Gamma_4^- ([111])$ ($R_4^+,\Gamma_4^- (d)$).
Inset: zoomed view of the magnetic energies of 
c-R3c (4) distortion.
}
\label{ediffdistbfmofig}
\end{center}
\end{figure}

Let us now look at the results for the structural energetics of the BiFeO$_3$-BiMnO$_3$ checkerboard.
In Fig.~\ref{ediffdistbfmofig}, we present 
the energies for four different types of structural distortions.
These types of distortions show the lowest energies among 
a larger set of structures that we explored.~\cite{Palova09}
Notice that the variation of the structural energy is much larger than 
that of the magnetic energy of the checkerboard.

Not surprisingly, the $R_4^+ ([111])$ and $\Gamma_4^- ([111])$ (R3c) type of structural distortion
is energetically the most favorable; it is
the BiFeO$_3$ ground state and the BiMnO$_3$ lowest energy structural 
distortion.
The R3c symmetry is now broken due to pillar cation ordering 
and the space group of the checkerboard ground state becomes $P1$;
we use the notation c-R3c, where c designates ``checkerboard'', as a reminder of the origin of the distortions.
As shown in the inset of Fig.~\ref{ediffdistbfmofig}, the two lowest magnetic states G-AFM and FeAFMMnFM, 
are only $2 meV/f.u.$ apart.
The ground state of the checkerboard
has the FeAFMMnFM magnetic order, where Fe spins are ordered antiferromagnetically (AFM)
along the Fe pillars, Mn spins are ordered ferromagnetically (FM) along the Mn pillars, reflecting
``AFM'' and ``FM'' nature of the parent BiFeO$_3$ and BiMnO$_3$ compounds respectively. 
A total magnetic moment $3.7\mu_B$ per Fe -Mn pair results from manganese chains. 
The FeAFMMnFM ground state is insulating with energy gap $0.88 eV$,
and we calculate a value of the polarization 
$62\mu C/cm^2$ pointing in the $[0.85,0.85,1]$ direction.
The ground state of the checkerboard is multiferroic, being ferroelectric and ferrimagnetic.

In particular we want to relate and
contrast the properties of the 
BiFeO$_3$-BiMnO$_3$ checkerboard to those of its two bulk constituent
materials;
we recall that 
BiFeO$_3$ is polar and antiferromagnetic,
while 
BiMnO$_3$ is non-polar and ferromagnetic, and we have found
that the checkerboard assumes the desired ferromagnetic-ferroelectric
properties of each leading to a multiferroic ground-state whose
magnetic behavior is structurally sensitive.  We attribute this
behavior to the development of a quasi-degenerate manifold of
magnetic states in the checkerboard, in contrast to the gapped
states in the layered superlattice and in the bulk; this can
be understood in terms of frustration of the cations inherent in the
checkerboard geometry.
Since bulk BiFeO$_3$ is known to be G-AFM, and bulk BiMnO$_3$ FM, 
the Fe-Fe and
Mn-Mn interactions tend to be AFM and FM-like respectively.
In the layered superlattice, each Fe(Mn) atom has 
four Fe(Mn) and only two 
Mn(Fe) nearest neighbours, 
so that the Fe/Mn layers prefer to be AFM/FM,
leading to minor frustration between the minority of mixed Fe-Mn bonds.
The FeAFMMnFM layered ground state is much more preferable and lower in energy
than any other magnetic state.
In the checkerboard there are more frustrated bonds per each cation,
and therefore more weight is given to the mixed Fe-Mn bonds and 
various magnetic states compete energetically.

We study the sensitivity of the closely spaced magnetic levels 
in the checkerboard to a structural distortion.
As we tune the system from the checkerboard c-R3c ground state to 
c-I4cm state with $R_4^+ (y)\&\Gamma_4^- (y)$ distortions,
either the FeAFMMnFM (filled diamond)
or the G-AFM (open circle) lowest state
is favored.
Switching between these two magnetic states occurs 
as we tune the system to other structural distortions (see Fig.~\ref{ediffdistbfmofig}).
It is the competition between these two magnetic types that allows switching 
between nonzero and zero magnetization;
the magnetostructural effect leads to the possibility of a structurally-driven 
magnetic transition in the checkerboard.
This could be realized, for example, by imposing expitaxial strain constraints.

In summary, we present a magnetostructural effect in the atomic-scale
checkerboard BiFeO$_3$-BiMnO$_3$, which is not present in either bulk
or in layered structures of these two materials.  Furthermore,
unlike its parent compounds, the checkerboard has a multiferroic ground
state with a nonzero magnetization and polarization; this is a new example
of a nanocomposite with properties that can be designed. We note that
this behavior is due to the magnetic frustration in this system inherent
to the checkerboard geometry; as a result the magnetic states are quasi-degenerate and can be tuned by small perturbations including strain. We remark 
that our first principles calculations do not include spin-orbit coupling
which is known to lead to weak ferromagnetism in bulk BiFeO$_3$.~\cite{Ederer05}
We expect that such corrections will not change our results
fundamentally, but this is certainly worth pursuing in future work.
We would also plan to investigate similar checkerboards on longer 
length-scales to make more direct contact with the possibility of 
future experiments.

We thank V. R. Cooper, M. Dawber, C.-J. Eklund, C. Fennie, A. Malashevich, 
M. Marsman and D. Vanderbilt for helpful discussions. This work 
was supported in part by NSF MRSEC DMR-0820404, NSF NIRT-ECS-0608842
and by the US Army Research Office through  MURI-DAAD 19-01-1-0517.

\def\refname{Bibliography}

\end{document}